\documentclass[a4paper,10pt]{article}
\usepackage{enumerate}
\usepackage{amssymb, amsmath, amsfonts, amsthm}
\usepackage[utf8]{inputenc} 
\usepackage{float}
\usepackage{graphicx}
\usepackage{flushend}
\usepackage{multicol}
\usepackage[left=1.7cm,top=2.2cm,right=1.7cm,bottom=2.5cm]{geometry}

\usepackage{multirow}
\usepackage{slashed}

\begin{document}

\begin{center}
\Large{\textbf{Weyl spinors and the helicity formalism}}
\normalsize
\\ \vspace{0.5cm}
J. L. Díaz Cruz, Bryan Larios, Oscar Meza Aldama, Jonathan Reyes Pérez\\
\it{\normalsize Facultad de Ciencias F\'isico-Matem\'aticas, Benem\'erita Universidad Aut\'onoma de Puebla, Av. San Claudio y 18 Sur, C. U. 72570 Puebla, M\'exico}
\end{center}
\small
\noindent

\noindent
\\

\noindent

\large
\noindent

\smallskip
\normalsize
\indent \textbf{Abstract.} In this work we give a review of the original formulation of the relativistic wave equation for particles with spin one-half. Traditionally (\textit{à la Dirac}), it's proposed that the ``square root'' of the Klein-Gordon (K-G) equation involves a 4 component (Dirac) spinor and in the non-relativistic limit it can be written as 2 equations for two 2 component spinors. On the other hand, there exists Weyl's formalism, in which one works from the beginning with 2 component Weyl spinors, which are the fundamental objects of the helicity formalism. In this work we rederive Weyl's equations directly, starting from K-G equation. We also obtain the electromagnetic interaction through minimal coupling and we get the interaction with the magnetic moment. As an example of the use of that formalism, we calculate Compton scattering using the helicity methods. 

\section{Introduction}
One of the cornerstones of contemporary physics is quantum mechanics, thanks to which great advances in the comprehension of nature at the atomic, and even subatomic level have been achieved. On the other hand, its applications have given place to a whole new technological revolution. Thus, the study of quantum physics is part of our scientific culture. \newline
\indent We should remember that the first formulation of (ondulatory) Quantum Mechanics was based in the time-dependent Schr\"odinger equation:
\begin{equation}
	i\hbar\frac{\partial}{\partial t}\Psi = H\Psi ,
	\label{schrodinger_eq_t}
\end{equation}
where $H$ represents the hermitian operator that defines the total energy of the system, and the wave function $\Psi$ depends on both position and time: $\Psi =\Psi (\vec{r},t)$. For a free particle, the hamiltonian has the form $H=-\frac{\hbar^2}{2m}\nabla^2$. Applying the method of separation of variables, i.e., $\Psi(\vec{r},t)=\phi(t)\psi(\vec{r})$, we obtain \cite{griffiths_QM}
\begin{equation}
	\label{schrodinger_eq}
	-\frac{\hbar^2}{2m}\nabla^2\psi = E \psi ,
\end{equation}
where now $\psi(\vec{r})$ is a function of the spatial coordinates only. The solutions to equation \eqref{schrodinger_eq_t} are of the form
\begin{equation}
	\label{e1}
	\Psi = \psi \exp(-iEt / \hbar),
\end{equation}
where $E$ is the corresponding eigenvalue of $H$. With this expression we obtain a positive-definite probability density which, together with its probability current, satisifies the continuity equation. Actually, the condition $H=H^\dagger$ and having a first temporal derivative only ensure this property. \newline
\indent After the success of quantum mechanics based on equation \eqref{schrodinger_eq} to describe the atom, it was proposed to obtain a relativistic wave equation; such an equation is known (in honor to Oscar Klein and Walter Gordon\footnote{Although Schr\"odinger himself proposed it.}) as Klein-Gordon (K-G) equation, and it's of the form\footnote{From now on, we'll use natural units, i.e., $\hbar =c=1$.}:
\begin{equation}
	\label{intr_kg_eq}
	(\square + m^2)\psi =0 ,
\end{equation}
where $\square = \partial^\mu \partial_\mu =  \dfrac{{\partial}^2}{{\partial t^2}} - \nabla^2 $ is the so-called \emph{d'Alembertian} operator. This equation is relativistically invariant and it's used to describe  charged or neutral spinless relativistic particles, but its probability density is not positive-definite and it admits negative-energy solutions (see \cite{ryder}); also, it doesn't describe the elctrons correctly, because they have spin one half. \newline
\indent In 1928, Dirac proposed a relativistic expression in which both spatial and temporal derivatives appeared at first order, instead of the second order in which they appear in the d'Alembertian operator of equation \eqref{intr_kg_eq}, hoping somehow to solve the problem of the non-definite-positive probability density. To obtain a positive probability density, Dirac proposed taking the ``square root'' of \eqref{intr_kg_eq}, using objects (Dirac matrices) that satisfy a certain anticommutation relation, and arrived at the following expression:
\begin{equation}
	i\frac{\partial\psi}{\partial t}=(-i\vec{\alpha}\cdot\nabla+\beta m)\psi ,
\end{equation}
that we can rewrite in the following covariant form:
\begin{equation}
	\label{intr_dirac_eq}
	(i\gamma^\mu \partial_\mu - m ) \psi = 0 ,
\end{equation}
\noindent
which is known as Dirac equation, and where $\psi$ is now an object called $spinor$ (with 4 components). Furthermore, using minimal substitution $\partial_\mu$ $\mapsto$ $D_\mu = \partial_\mu - ieA_\mu$, it's possible to describe the interactions of electrons with the electromagnetic potentials. \newline
\indent Equation \eqref{intr_dirac_eq} has two independent solutions (four, if we take into account the spin orientation): one with positive charge that describes the electron, and another one corresponding to a particle with positive charge, which is now called \emph{positron}. The only particles known in 1928 were the electrons, the protons and the photons, so Dirac thought that the positive charge solution should correspond to the proton, until in 1932 Carl Anderson discovered, while studying cosmic rays in a cloud chamber, that some trajectories deviated with the same magnitude as the electron but in opposite direction. He had discovered the first antiparticle: the positron. \newline
\indent In that same year James Chadwick experimentally proved the existence of neutrons, which were necessary to understand the stability of the atomic nucleus, but it wasn't understood how did it hold together. In 1935 Hideki Yukawa reasoned that there should exist another force, more powerful than the electromagnetic, which kept together the protons although they repel each other. He proposed the existence of a field analogous to the electromagnetic one that produced massive bosons, responsible for the strong nuclear force, and from the short range of those interactions he predicted that their mass should be of the order of $150$ MeV. In 1947, english physicist Cecil Powell (in an experiment with cosmic rays) found a particle with the exact properties predicted by Yukawa, which he denominated \emph{pion}. \newline
\indent At the end of the nineteenth century, Rutherford discovered that the uranium emits two types of radiation: alpha radiation, which is helium nuclei (two protons and two neutrons), and beta radiation, which is electrons. Later on, Rutherford and Soddy demonstrated that for radioactive atoms with unstable nuclei, by emitting an electron ($\beta$ radiation), a neutron converts into a proton, giving as a result an atom of a different element. On the other hand, when a neutron is free, it disintegrates to form an electron and a proton ($\beta$ disintegration):
\begin{equation}\label{e2}
	n \longrightarrow p + e^{-} ,
\end{equation}
but the sum of the energies of the proton and the electron produced in this process was less than the one of the neutron. Some experiments were conducted later for neutrons at rest and it was demonstrated that the magnitudes of the momenta of the emitted electrons had distinct values, but for this process the energy conservation law is
\begin{equation}\label{e2}
 c^{2} M_n = (M_e c^{4} + P_e^{2} c^{2})^{\frac{1}{2}} + (M_p c^{4} + P_e^{2} c^{2})^{\frac{1}{2}},
\end{equation}
where $M_n$, $M_e$, $M_p$ are the neutron, electron and proton masses, respectively, and $P_e^{2}$ amd $P_p^{2}$ are the magnitudes of the momenta of the electron and the proton. Because the neutrino is at rest, $\vec{P}_n=0=\vec{P}_p+\vec{P}_e$, which implies $|\vec{P}_p|=|\vec{P}_e|$, and then we have a unique solution for $P_e$, fixed by the masses of the particles. This represented a contradiction to the experiments, on which it was found that there is a distribution for the electron energy. Some even thought that the energy conservation law wasn't valid for some processes between particles, until in 1930 Wolfgang Pauli suggested the existence of a new particle with zero mass and zero electric charge; therefore the process should be
\begin{equation}\label{e2}
 n \longrightarrow p + e^{-} + \overline{\nu_e} .
\end{equation}
Three years later Enrico Fermi, influenced by Pauli's idea, introduced the weak nuclear force to explain this phenomenon and proposed that the neutron becomes a proton, emitting a charged boson ($W^-$); Fermi named this new particle ($\overline{\nu_e}$) \emph{neutrino}, currently it's called electron antineutrino. \newline
\indent To describe particles and antiparticles of spin one-half, in 1929 Hermann Weyl proposed a pair of coupled equations that bear his name and, for the massless case (neutrinos, for example), the equations decouple and take the form
\begin{equation}\label{intr_weyl_1}
 i \dfrac{\partial}{\partial t}\phi = i \overrightarrow{\sigma}\cdot\nabla\phi,
\end{equation}
\begin{equation}\label{intr_weyl_2}
 i \dfrac{\partial}{\partial t} \chi =-i \overrightarrow{\sigma} \cdot \nabla \chi,
\end{equation}
where $\phi$ and $\chi$ are objects called, respectively, left and right\footnote{The terminology comes from a certain representation of the complexification of the Lie algebra of the Lorentz group; see \cite{ryder}.} \emph{Weyl spinors} (2-component). A year before Weyl, Dirac had considered them starting from the 2 component version of his equation, but Pauli (see \cite{penrose}) didn't accept them because they are not invariant under parity, \newline
\indent In 1956, Tsung Dao Lee and Chen Ning Yang (\cite{lee_yang_1}, \cite{lee_yang_2}) proposed that parity is a symmetry of al processes caused by strong and electromagnetic interactions, but it's not in processes due to weak interactions such as, for example, $\beta$ disintegration. A year later Chien Shiung Wu \cite{wu} and her collabborators analyzed the $\beta$ disintegration of a Cobalt 60 neutron and experimentally proved that weak interactions violate parity: the reaction gave rise to a nickel nucleus and the emission of electrons and electronic antineutrinos. \newline
\indent In 1954, Chen Ning Yang and Robert Mills \cite{yang_mills} succeeded in constructing a field theory invariant non-abelian transformations. Then, in 1957 Julian Schwinger \cite{schwinger} took this theory and applied it to the weak nuclear force and to the electromagnetic force and realized that this forces had a similar magnitude, but this symmetry was broken because the gauge bosons $W^\pm$ of the weak force have a mass, whereas the gauge boson of the electromagnetic field (the photon) is massless. A year after, Sidney Bludman \cite{bludman} suggested that the weak nuclear force may be described through a local, non-abelian gauge theory and he introduced three particles, $W^{+}$, $W^{-}$ and $Z^{0}$; the $Z^{0}$ particle described weak interactions in which the electric charge doesn't play a role. On the other hand, in 1961 Sheldon Glashow (\cite{glashow_1}, \cite{glashow_2}), using Bludman's theory, created a model which included a triplet of vector bosons, carriers of the weak force, and a vector boson carrying the electromagnetic force; it was realized that the triplet and the singlet could combine in such a way that a new neutral massive particle ($Z^{0}$) would emerge, but the other one (photon) would remain massless. \newline
\indent Paul Dirac, along with Jordan, Heisenberg and Pauli, formulated the first quantum field theories, but there appeared infinities within them. The contributions of Tomonaga, Bethe, Dyson, Schwinger and Feynman (with his diagrams that allow us to study the electromagnetic interactions of objects such as electrons, positrons, quarks, etc.) improved the theories to be renormalizable, and quantum electrodynamics (QED) turned into the simplest quantum field theory that is invariant under gauge transformations. Later developments by Ward, Heisenberg, Salam, Wilson, Veltman and t'Hooft  led Yang-Mills theories to be renormalizable, and these are the base of modern particle physics. \newline
\indent Traditionally in QED we use Dirac's formalism (4-component) instead of Weyl's formalism (2-components). In this article we try to construct the complete theory of QED starting from the beginning with Weyl spinors. Currently, this program of trying to formulate quantum field theories (QFT) with definite helicity fields has been extended to the study of processes with a lot of particles in the final state and is one of the most active research areas in QFT. It is possible that in the future QFT will be studied from the beginning with this helicity formalism. \newline
\indent The content of this article is the following: In section \ref{sec:weyl_spinors} we obtain equations \eqref{intr_weyl_1} and \eqref{intr_weyl_2} starting from \eqref{intr_kg_eq}. In section \ref{sec:EM_weyl_coupling} we derive the equations of motion for Weyl fields with spatial and temporal derivatives and we verify that the system is invariant under parity and charge conjugation transformations. In section \ref{sec:fhelicidad} we introduce the notation used in the helicity formalism. Finally, in section \ref{sec:ComptonScatering} we use the helicity formalism to calculate the invariant amplitude of the electron-photon scattering, callled Compton scattering, at tree level and we will see the advantages that one has by using this formalism.

\section{Weyl spinors}
\label{sec:weyl_spinors}
In this section we obtain Weyl equations starting from K-G equation, and we prove some properties of 2-component spinors. K-G equation\footnote{Henceforth, unlike \eqref{intr_kg_eq}, we use the Minkowski metric tensor with signature $+2$, i.e., $(g_{\mu\nu})=\text{diag}(-1,+1,+1,+1)$.} is given by
\begin{equation}
	(-\square +m^2)\phi =0,
\end{equation}
which we rewrite as
\begin{equation}
	\left( -\frac{\partial^2}{\partial t^2}+\nabla^2\right)\phi =m^2\phi .
	\label{KG}
\end{equation}
The goal is to obtain a first order differential equation, so we can try to ``factorize'' the differential operator on the LHS of the previous equation. For this matter we propose to write it as follows
\begin{equation}
	-\frac{\partial^2}{\partial t^2}+\nabla^2 =\left( i\frac{\partial}{\partial t} +\vec{\alpha}\cdot\nabla\right)\left(i\frac{\partial}{\partial t} +\vec{\beta}\cdot\nabla\right) ,
	\label{factorization}
\end{equation}
where $\vec{\alpha}$ and $\vec{\beta}$ are constant (three-)vectors (independent of the spacetime coordinates, the field and its derivatives) which are as yet undetermined. Expanding the RHS of \eqref{factorization}:
\begin{equation}
	-\frac{\partial^2}{\partial t^2}+\nabla^2=-\frac{\partial^2}{\partial t^2}+i(\vec{\beta}+\vec{\alpha})\cdot\nabla\frac{\partial}{\partial t}+(\vec{\alpha}\cdot\nabla )(\vec{\beta}\cdot\nabla ) .
\end{equation}
For both sides to be equal, we must have
\begin{equation}
	\vec{\beta}=-\vec{\alpha}
	\label{minus_alpha}
\end{equation}
and
\begin{equation}
	(\vec{\alpha}\cdot\nabla )(\vec{\beta}\cdot\nabla ) =\nabla^2 ,
	\label{laplacian}
\end{equation}
which, using (\ref{minus_alpha}), implies
\begin{equation}
	\alpha_i^2=-1 .
	\label{alpha_square}
\end{equation}
For convenience, we define $\vec{\sigma}$ through the expression $\vec{\alpha}\equiv i\vec{\sigma}$; therefore, \eqref{alpha_square} is equivalent to
\begin{equation}
	\sigma_i^2=1 .
	\label{sigma_squared}
\end{equation}
\indent Now, we can rewrite equation (\ref{laplacian}), using the summation convention over repeated indices and the definition of $\vec{\sigma}$, as follows:
\begin{equation}
\begin{split}
	\nabla^2=\partial_i\partial_i &= \alpha_i\beta_j\partial_i\partial_j \\
	&= \frac{1}{2}(\alpha_i\beta_j+\alpha_j\beta_i)\partial_i\partial_j \\
	&= -\frac{1}{2}(\alpha_i\alpha_j+\alpha_j\alpha_i)\partial_i\partial_j \\
	&= \frac{1}{2}(\sigma_i\sigma_j+\sigma_j\sigma_i)\partial_i\partial_j .
\end{split}
\end{equation}
Therefore, the  components of $\vec{\sigma}$ should satisfy
\begin{equation}
	\sigma_i\sigma_j+\sigma_j\sigma_i\equiv\left\lbrace\sigma_i,\sigma_j\right\rbrace = 2\delta_{ij} .
	\label{anticommutator}
\end{equation}
No set of three real or complex numbers can fulfill equation (\ref{anticommutator}), but a set of matrices can satisfy such anticommutation relations. The ``canonical'' matrices that satisfy \eqref{anticommutator} are the \emph{Pauli matrices}:
\begin{equation}
	\sigma_1=\left(\begin{array}{cc}
	0 & 1 \\
	1 & 0
	\end{array}\right) \quad , \quad
	\sigma_2=\left(\begin{array}{cc}
	0 & -i \\
	i & 0
	\end{array}\right) \quad , \quad
	\sigma_3=\left(\begin{array}{cc}
	1 & 0 \\
	0 & -1
	\end{array}\right) .
	\label{pauli_matrices}
\end{equation}
Now that we know that $\vec{\sigma}$ is a vector of matrices, we should understand that there's a unit matrix multiplying the RHS of equations such as (\ref{sigma_squared}) and (\ref{anticommutator}). \newline
\indent Pauli matrices form a maximal set of anticommuting matrices, i.e., there does not exist a fourth matrix that anticommutes with all $\sigma_1$, $\sigma_2$ and $\sigma_3$, and whose square is the identity. To see this, first notice that $\sigma_3=-i\sigma_1\sigma_2$; indeed:
\begin{equation}
	-i\sigma_1\sigma_2=-i
	\left(\begin{array}{cc}
	0 & 1 \\
	1 & 0
	\end{array}\right)
	\left(\begin{array}{cc}
	0 & -i \\
	i & 0
	\end{array}\right) =-i
	\left(\begin{array}{cc}
	i & 0 \\
	0 & -i
	\end{array}\right) =
	\left(\begin{array}{cc}
	1 & 0 \\
	0 & -1
	\end{array}\right) =\sigma_3 .	
\end{equation}
Assuming there is a matrix $\sigma_4$ that anticommutes with the other three, we have
\begin{equation}
\begin{split}
	& \quad \quad \sigma_1\sigma_2\sigma_3\sigma_4=-\sigma_4\sigma_1\sigma_2\sigma_3 \\
	& \Rightarrow \sigma_1\sigma_2(-i\sigma_1\sigma_2)\sigma_4=-\sigma_4\sigma_1\sigma_2(-i\sigma_1\sigma_2) \\
	& \Rightarrow -\sigma_1\sigma_1\sigma_2\sigma_2\sigma_4=\sigma_4\sigma_1\sigma_2\sigma_1\sigma_2 \\
	& \Rightarrow -(\sigma_1)^2(\sigma_2)^2\sigma_4=\sigma_4(\sigma_1)^2(\sigma_2)^2 \\
	& \Rightarrow -\sigma_4=\sigma_4 \\
	& \Rightarrow \sigma_4=0 .
\end{split}
\end{equation}
Again, in the last line it's understood that $0$ represents the 2$\times$2 matrix whose entries are all zero. \newline
\indent Now we can rewrite \eqref{KG} as
\begin{equation}
	\left( i\frac{\partial}{\partial t} +i\vec{\sigma}\cdot\nabla\right)\left(i\frac{\partial}{\partial t} -i\vec{\sigma}\cdot\nabla\right)\phi =m^2\phi .
	\label{KG_factorized}
\end{equation}
The result of applying the differential operator $i\frac{\partial}{\partial t} -i\vec{\sigma}\cdot\nabla$ over $\phi$ will certainly be an object of the same type as $\phi$. Then we can write, without loss of generality,
\begin{equation}
	\left( i\frac{\partial}{\partial t} -i\vec{\sigma}\cdot\nabla\right)\phi =m\chi ,
	\label{weyl_1}
\end{equation}
which turns (\ref{KG_factorized}) into
\begin{equation}
	\left( i\frac{\partial}{\partial t} +i\vec{\sigma}\cdot\nabla\right)\chi =m\phi .
		\label{weyl_2}
\end{equation}
Rearranging \eqref{weyl_1} and \eqref{weyl_2}, we obtain respectively
\begin{equation}
	i\dfrac{\partial}{\partial t}\phi=i\vec{\sigma}\cdot\nabla\phi+m\chi
	\label{weyl_eq_1}
\end{equation}
and
\begin{equation}
	i\dfrac{\partial}{\partial t}\chi=-i\vec{\sigma}\cdot\nabla\chi+m\phi .
	\label{weyl_eq_2}
\end{equation}
Equations \eqref{weyl_eq_1} and \eqref{weyl_eq_2} are called (coupled) \emph{Weyl equations}. Each one of the matrices $\sigma_i$ is a 2$\times$2 matrix which acts on $\phi$ and $\chi$, so these must be two-component objects, which are called left and right Weyl spinors, respectively. Physically, they represent spin $\frac{1}{2}$ particles and antiparticles. \newline
\indent The zero mass case is particularly interesting; in this limit Weyl equations decouple and, using the canonical substitution $E\leftrightarrow i\frac{\partial}{\partial t}$ and $\vec{p}\leftrightarrow -i\nabla$, turn into
\begin{equation}
\begin{array}{c}
	E\phi =-\vec{\sigma}\cdot\vec{p}\phi , \\
	E\chi =\vec{\sigma}\cdot\vec{p}\chi .
\end{array}
\end{equation}
Remembering the relativity relationship $E^2-\vec{p}^2=m^2$, in the massless case we have $E=|\vec{p}|$, therefore,
\begin{equation}
\begin{array}{c}
	\vec{\sigma}\cdot\hat{p}\phi =-\phi , \\
	\vec{\sigma}\cdot\hat{p}\chi =\chi ,
\end{array}
	\label{massless_weyl_eqs}
\end{equation}
where $\hat{p}\equiv\frac{\vec{p}}{|\vec{p}|}$. From quantum mechanics, we know (see, for instance, \cite{griffiths_QM} or \cite{cohen}) that Pauli matrices, apart from a certain multiplicative constant, represent the spin $\frac{1}{2}$ operators, then $\vec{\sigma}\cdot\hat{p}$ is the operator that gives us the component of the spin in the direction of the linear momentum. This quantity (the value of the projection of a particle's spin along the direction of its momentum) is known as \emph{helicity}. Formally, we define the helicity operator as $h\equiv\vec{\sigma}\cdot\hat{p}$, so the equations \eqref{massless_weyl_eqs} tell us that the Weyl spinors $\phi$ y $\chi$ are helicity eigenstates, with eigenvalues $-1$ and $+1$, respectively. Some authors \cite{peskin} define a right particle to be one such that its helicity is $+\frac{1}{2}$, and a left particle as one with helicity $-\frac{1}{2}$. \newline
\indent As we said before, Weyl spinors have two components. A very used convention (see \cite{torres}, \cite{dreiner}) is to use dotted superindices ($\dot{a}$, $\dot{b}$, $... =\dot{1}, \dot{2}$) to label the components of the right spinor $\chi$ and undotted subindices ($a$, $b$, $...=1,2$) for the components of the left spinor $\phi$, and represent them with column vectors, i.e.,
\begin{equation}
	\phi_a =\left(\begin{array}{c} \phi_1 \\ \phi_2 \end{array}\right)
	\quad , \quad
	\chi^{\dot{a}} =\left(\begin{array}{c} \chi^{\dot{1}} \\ \chi^{\dot{2}} \end{array}\right) ,
	\label{spinor_columns}
\end{equation}
where $\phi_1$, $\phi_2$, $\chi^{\dot{1}}$ and $\chi^{\dot{2}}$ are, in general, complex numbers (which commute among themselves). The two-index Levi-Civita symbol is used to raise or lower spinor indices:
\begin{equation}
\begin{array}{c}
	\phi^a\equiv\epsilon^{ab}\phi_b , \\
	\chi_{\dot{a}}\equiv\epsilon_{\dot{a}\dot{b}}\chi^{\dot{b}} ,
	\label{raise_lower_indices}
\end{array}
\end{equation}
where $\epsilon^{12}=\epsilon^{\dot{1}\dot{2}}=\epsilon_{21}=\epsilon_{\dot{2}\dot{1}}\equiv +1$. Note that, using the antisymmetry of $\epsilon_{ab}$,
\begin{equation}
	\phi^a\psi_a=\phi^a\epsilon_{ab}\psi^{b}=-\epsilon_{ba}\phi^a\psi^b=-\phi_a\psi^a ,
	\label{spinor_product}
\end{equation}
and therefore
\begin{equation}
	\phi^a\phi_a=0 .
	\label{spinor_null}
\end{equation}
In analogy with \eqref{spinor_columns}, the spinors $\phi^a$ and $\chi_{\dot{a}}$ are usually represented with row vectors:
\begin{equation}
	\phi^a=(\phi^1 , \phi^2) \quad , \quad \chi_{\dot{a}}=(\chi_{\dot{1}} , \chi_{\dot{2}}) .
\end{equation}
\indent The components $\phi_a$ and $\chi_{\dot{a}}$ of the Weyl spinors are not independent, as we shall see. First note that the definition \eqref{raise_lower_indices} implies that, numerically,
\begin{equation}
\begin{aligned}
	\phi^1&=\phi_2 , \\
	\phi^2&=-\phi_1 .
\end{aligned}
	\label{up_down_indices}
\end{equation}
Furthermore, Weyl ``left equation'', in matrix form, is
\begin{equation}
	\left(\begin{array}{c}
		E\phi_1 \\ E\phi_2
	\end{array}\right) =-
	\left(\begin{array}{cc}
		p_3 & p_1-ip_2 \\
		p_1+ip_2 & -p_3
	\end{array}\right)
	\left(\begin{array}{c}
		\phi_1 \\ \phi_2
	\end{array}\right) ,
\end{equation}
which is equivalent, equating components, using \eqref{up_down_indices} and taking complex conjugate, to
\begin{subequations}
	\begin{equation}
		E\phi^{2\ast}=-p_3\phi^{2\ast}+(p_1+ip_2)\phi^{1\ast} ,
		\label{left_weyl_1}
	\end{equation}
	\begin{equation}
		E\phi^{1\ast}=(p_1-ip_2)\phi^{2\ast}+p_3\phi^{1\ast} .
		\label{left_weyl_2}
	\end{equation}
\end{subequations}
On the other hand, Weyl ``right equation'' is given by:
\begin{subequations}
	\begin{equation}
		E\chi^{\dot{1}}=p_3\chi^{\dot{1}}+(p_1-ip_2)\chi^{\dot{2}} ,
		\label{right_weyl_1}
	\end{equation}
	\begin{equation}
		E\chi^{\dot{2}}=(p_1+ip_2)\chi^{\dot{1}}-p_3\chi^{\dot{2}} .
		\label{right_weyl_2}
	\end{equation}
\end{subequations}
You can see that \eqref{left_weyl_1} and \eqref{left_weyl_2} have exactly the same form that \eqref{right_weyl_2} and \eqref{right_weyl_1}, respectively, just interchanging $\chi^{\dot{1}}$ by $\phi^{1\ast}$ and $\chi^{\dot{2}}$ by $\phi^{2\ast}$. Thus, we conclude that\footnote{Note that, in the argument given, we assumed that the momentum $\vec{p}$ (and the energy $E$) is real. In some cases it's useful to work with complexified momenta; see \cite{huang}. If the momentum $p$ is complex, relations \eqref{dot_undot_rel} are invalid.}
\begin{equation}
\begin{aligned}
	\chi^{\dot{a}} &= \phi^{a\ast} , \\
	\chi_{\dot{a}} &= \phi_a^\ast .
\end{aligned}
	\label{dot_undot_rel}
\end{equation}

\section{Coupling of Weyl fields to the electromagnetic field}
\label{sec:EM_weyl_coupling}
The lagrangian (density) that describes two Weyl spinor fields can be written as (see \cite{srednicki}):
\begin{equation}
	\mathcal{L}=i\chi^\dagger\bar{\sigma}^\mu\partial_\mu\chi +i\phi^\dagger\bar{\sigma}^\mu\partial_\mu\phi -m\chi\phi -m\chi^\dagger\phi^\dagger ,
\end{equation}
where $\bar{\sigma}^\mu =(-I, \vec{\sigma})$. Using Euler-Lagrange equations, we obtain the equations of motion (EOM)
\begin{equation}
	m\chi -i\sigma^\mu\partial_\mu\phi^\dagger =0 ,
\end{equation}
\begin{equation}
	-i\bar{\sigma}^\mu\partial_\mu\chi +m\phi^\dagger =0 ,
	\label{EL_EOM_chi}
\end{equation}
where $\sigma^\mu =(I, \vec{\sigma})$. In the case $m=0$ (for simplicity), using the minimal substitution $\partial_\mu\rightarrow D_\mu =\partial_\mu -ieA_\mu$ ($A_\mu$ is the electromagnetic four-potential: $A_\mu=(\varphi,\vec{A})$, where $\varphi$ is the electric scalar potential and $\vec{A}$ is the magnetic vector potential) in the previous equations, we get
\begin{equation}
	-i\sigma^\mu D_\mu\phi^\dagger =0 ,
	\label{EOM_phi}
\end{equation}
\begin{equation}
	-i\bar{\sigma}^\mu D_\mu\chi =0 .
	\label{EOM_chi}
\end{equation}
\indent The goal is to arrive, starting from \eqref{EOM_phi} and \eqref{EOM_chi}, to EOM with second order spatial and temporal derivatives. Multiplying \eqref{EOM_phi} by the operator $i\bar{\sigma}^\nu D_\nu$, we obtain
\begin{equation}
	\bar{\sigma}^\nu\sigma^\mu D_\nu D_\mu\phi^\dagger =0 .
	\label{sigma_kg}
\end{equation}
Then, using the following relationships (see \cite{dreiner}):
\begin{equation}
	\bar{\sigma}^\nu\sigma^\mu =g^{\nu\mu}-2i\bar{\sigma}^{\nu\mu} ,
	\label{sigmas}
\end{equation}
\begin{equation}
	\bar{\sigma}^{\nu\mu}=-\bar{\sigma}^{\mu\nu} ,
	\label{as1}
\end{equation}
\begin{equation}
	\sigma^{\nu\mu}=-\sigma^{\mu\nu} ,
	\label{as2}
\end{equation}
where
\begin{gather}
	\sigma^{\mu\nu}\equiv\frac{i}{4}(\sigma^\mu\bar{\sigma}^\nu-\sigma^\nu\bar{\sigma}^\mu) , \\
	\bar{\sigma}^{\mu\nu}\equiv\frac{i}{4}(\bar{\sigma}^\mu\sigma^\nu-\bar{\sigma}^\nu\sigma^\mu) ,
\end{gather}
we get
\begin{equation}
	\left( D_\mu D^\mu -2i\bar{\sigma}^{\nu\mu} D_\nu D_\mu\right)\phi^\dagger =0 .
	\label{order_2_EOM}
\end{equation}
The second term of \eqref{order_2_EOM} must be analyzed a little further; using \eqref{as1} we see that
\begin{equation}
\begin{split}
	\bar{\sigma}^{\nu\mu} (\partial_\nu -ieA_\nu )(\partial_\mu -ieA_\mu )\phi^\dagger & = -ie\bar{\sigma}^{\nu\mu}\partial_\nu A_\mu\phi^\dagger \\
	& = -\frac{ie}{2}(\bar{\sigma}^{\nu\mu}\partial_\nu A_\mu +\bar{\sigma}^{\mu\nu}\partial_\mu A_\nu )\phi^\dagger \\
	& = -\frac{ie}{2}\bar{\sigma}^{\nu\mu}F_{\nu\mu} .
\end{split}
\end{equation}
We've reduced $\bar{\sigma}^{\nu\mu} D_\nu D_\mu =-\frac{ie}{2}\bar{\sigma}^{\nu\mu} F_{\nu\mu}$ because we have
\begin{equation}
	\bar{\sigma}^{\nu\mu}\partial_\nu\partial_\mu\phi^\dagger =0 ,
\end{equation}
\begin{equation}
	\bar{\sigma}^{\nu\mu} (A_\mu\partial_\nu +A_\nu\partial_\mu )\phi^\dagger =0
\end{equation}
and
\begin{equation}
	\bar{\sigma}^{\nu\mu}A_\nu A_\mu\phi^\dagger =0 .
\end{equation}
Finally, \eqref{sigma_kg} is
\begin{equation}
	(D_\mu D^\mu -e\bar{\sigma}^{\nu\mu} F_{\nu\mu} )\phi^\dagger =0.
	\label{FEOM_phi}
\end{equation}
With the same procedure we can find that \eqref{EL_EOM_chi} is
\begin{equation}
	(D_\mu D^\mu -e\sigma^{\nu\mu} F_{\nu\mu} )\chi =0.
	\label{FEOM_chi}
\end{equation}
\indent To understand the coupling of a spin $s=\frac{1}{2}$ to an external electromagnetic field, we must express \eqref{FEOM_phi} and \eqref{FEOM_chi} in a way such that the fields $\vec{E} (\vec{r} ,t)$ and $\vec{B} (\vec{r} ,t)$ appear coupled explicitly to the Pauli matrices. From the second term of \eqref{FEOM_phi} we have
\begin{equation}
\begin{split}
	\bar{\sigma}^{\nu\mu} F_{\nu\mu} &=\bar{\sigma}^{00} F_{00} +\bar{\sigma}^{0j} F_{0j} +\bar{\sigma}^{j0} F_{j0} +\bar{\sigma}^{ij} F_{ij} \\
	&= \frac{1}{2}i\sigma^jF_{0j} -\frac{1}{2}i\sigma^jF_{j0} +\frac{1}{2}\epsilon^{ijk}\sigma_k F_{ij} \\
	&= \frac{1}{2}i\vec{\sigma}\cdot\vec{E} +\frac{1}{2}i\vec{\sigma}\cdot\vec{E} -\vec{\sigma}\cdot\vec{B} \\
	&= -\vec{\sigma}\cdot (\vec{B} -i\vec{E} ) ,
\end{split}
	\label{sigma_E_B}
\end{equation}
where we've used the following identities (see \cite{dreiner}):
\begin{equation}
	\sigma^{ij} =\bar{\sigma}^{ij} =\frac{1}{2}\epsilon^{ijk}\sigma^k
\end{equation}
and
\begin{equation}
	\sigma^{i0} =-\sigma^{0i} =-\bar{\sigma}^{i0} =\bar{\sigma}^{0i} =\frac{1}{2} i\sigma^i .
\end{equation}
On the other hand, we know that the components of the fields $\vec{E}=(E_1,E_2,E_3)$ and $\vec{B}=(B_1,B_2,B_3)$ are related to the Faraday tensor through
\begin{equation}
	B^i=-\frac{1}{2}\epsilon^{ijk} F_{jk}
\end{equation}
and
\begin{equation}
	E^i=-F^{0i}=F^{i0} ,
\end{equation}
where
\begin{equation}
	(F^{\mu\nu})= \left(\begin{array}{cccc}
	0 & -E^1 & -E^2 & -E^3 \\
	E^1 & 0 & -B^3 & B^2 \\
	E^2 & B^3 & 0 & B^1 \\
	E^3 & -B^2 & B^1 & 0
	\end{array}\right) .
\end{equation}
\indent Finally, using \eqref{sigma_E_B}, equation \eqref{FEOM_phi} takes the form
\begin{equation}
	(D_\mu D^\mu +e\vec{\sigma}\cdot (\vec{B} -i\vec{E}))\phi^\dagger =0 .
	\label{FEOM_phi_EB}
\end{equation}
In a similar way, for \eqref{FEOM_chi} we find
\begin{equation}
	(D^\mu D_\mu +e\vec{\sigma}\cdot (\vec{B} +i\vec{E} ))\chi =0 .
	\label{FEOM_chi_EB}
\end{equation}
Now that we have the EOM for two-component spinor fields, we can verify that the system has C and P symmetry.

\subsection{Parity}
We know that $\vec{E}$ is a polar vector and $\vec{B}$ is a pseudovector (or axial vector), i.e., under a parity transformation ($\vec{r}\rightarrow \vec{r}^P=-\vec{r}$), $\vec{E}\rightarrow -\vec{E}$ and $\vec{B}\rightarrow\vec{B}$. If we take \eqref{FEOM_chi_EB} and apply a parity transformation, we obtain
\begin{equation}
	(D^\mu D_\mu +e\vec{\sigma}\cdot (\vec{B}^P +i\vec{E}^P)) \chi^P =0
\end{equation}
\begin{equation}
	(D^\mu D_\mu +e\vec{\sigma}\cdot (\vec{B} -i\vec{E} ))\chi^P=0 .
\end{equation}
We see that, if we define $\chi^P=\phi^\dagger$, then equation \eqref{FEOM_chi_EB}, under parity transformation, is exactly \eqref{FEOM_phi_EB}. Conversely, by the same transformation, \eqref{FEOM_phi_EB} transforms into \eqref{FEOM_chi_EB}. Therefore the system is invariant under parity transformations.

\subsection{Charge conjugation}
If we take equation \eqref{FEOM_phi_EB} and perform charge conjugation, i.e.,
\begin{equation}
\begin{array}{c}
	e\rightarrow e^C=-e , \\
	\rho (\vec{r})\rightarrow\rho^C(\vec{r})=-\rho (\vec{r}) , \\
	V (\vec{r})\rightarrow V^C(\vec{r})=-V (\vec{r}) , \\
	\vec{A} (\vec{r})\rightarrow \vec{A}^C(\vec{r})=-\vec{A} (\vec{r}) ,
\end{array}
\end{equation}
we see that
\begin{equation}
	(D_\mu D^\mu +e^C\vec{\sigma}\cdot (\vec{B}^C-i\vec{E}^C))\phi^{\dagger C} =0
\end{equation}
\begin{equation}
	(D_\mu D^\mu -e\vec{\sigma}\cdot (-\vec{B}+i\vec{E}))\phi^{\dagger C} =0
\end{equation}
\begin{equation}
	(D_\mu D^\mu +e\vec{\sigma}\cdot (\vec{B}-i\vec{E}))\phi^{\dagger C} =0 .
\end{equation}
If we define $\phi^{\dagger C} =\phi^\dagger$, we see that \eqref{FEOM_phi_EB} is invariant under charge conjugation. Analogously for \eqref{FEOM_chi_EB}.

\section{Basic features of the helicity formalism (\textbf{HF})}
\label{sec:fhelicidad}
Traditionally, the basic processes of QED are handled in the 4-component spinor formalism, however, recently \cite{huang} the many advantages of working with helicity methods in the 2-component spinor formalism have been discovered. In this section we will present the aspects of QED in the helicity formalism. \newline
\indent Let $p$ and $k$ be two four-momenta, and let $\phi$ and $\kappa$ the corresponding spinors, i.e., $\phi$ is the Weyl spinor that represents a massless particle of spin one-half (arbitrary orientation) and with four-momentum $p$, and something similar for $\kappa$ and $k$. We define the notation
\begin{equation}
	[pk]\equiv[p|^a|k]_a\equiv\phi^a\kappa_a.
\end{equation}

(That is, we've defined $[p|^a\equiv\phi^a$ and $|k]_a\equiv\kappa_a$, and the notation $[pk]$ is simply an abbreviation for the contraction of undotted spinor indices.) Due to \eqref{spinor_product} and \eqref{spinor_null}, we have $[kp]=-[pk]$ and $[pp]=0$. For right spinors, in turn, we define
\begin{equation}
	\langle pk\rangle\equiv\langle p|_{\dot{a}}|k\rangle^{\dot{a}}\equiv\phi_{\dot{a}}\kappa^{\dot{a}}
\end{equation}
(here, we've defined $\langle p|_{\dot{a}}\equiv\phi_{\dot{a}}$ and $|k\rangle^{\dot{a}}\equiv\kappa^{\dot{a}}$, and $\langle pk \rangle$ is an abbreviation for dotted spinor indices), which implies $\langle kp\rangle = -\langle pk\rangle$ and $\langle pp \rangle =0$. Using \eqref{dot_undot_rel}, we have
\begin{equation}
	\langle pk \rangle =\phi_{\dot{a}}\kappa^{\dot{a}} = (\phi_a\kappa^a)^\ast = (\kappa^a\phi_a)^\ast = [kp]^\ast .
\end{equation}
\indent In the \textbf{HF} one works with two component spinors that commute (also known in the literature as twistors). We'll see that in the ultraenergetic limit or, equivalently, for $s$, $|t|$ and $|u|\gg m^2$ \footnote{$s=(p_1+p_2)^2=(p_3+p_4)^2$, $t=(p_1-p_3)^2=(p_2-p_4)^2$ and $u=(p_1-p_4)^2=(p_2-p_3)$ are the kinematical variables, invariant under Lorentz transformations, known as Mandelstam variables, in honor to the Southafrican physicist Stanley Mandelstam; $p_1,\,p_2$ are the incoming 4-momenta and $p_3,\,p_4$ are the outgoing 4-momenta.}, the calculation gets much simpler than using the conventional methods with 4-component spinors. \newline
\indent In order to be able to do the calculation in the \textbf{HF} it is convenient to use the new notation, that basically consists in writing the components of the usual spinors $u$ and $v$, of 4 components, in terms of twistors:
\begin{equation}
\label{eq:notationtwistors}
\begin{split}
	u_-(p)=v_+(p)=\left(\begin{array}{c} |p]_a \\ 0\end{array}\right) , &\qquad u_+(p)=v_-(p)=\left(\begin{array}{c} 0 \\ |p\rangle^{\dot{a}}\end{array}\right) , \\
	\overline{u}_-(p)=\overline{v}_+(p)=(0,\langle p|_{\dot{a}}) , &\qquad \overline{u}_+(p)=\overline{v}_-(p)=([p|^a,0) ,
\end{split}
\end{equation}
where the product between these brackets obeys
\begin{equation}
\label{eq:braketstwistors}
\begin{split}
\left[k \right| \left| p \right] & = [k\, p],
\qquad
\left\langle k \right|\left|p \right\rangle = \langle k\, p \rangle,
\\
\left[k \right|\left| p \right\rangle & = 0,
\qquad
\hspace{0.4cm}\left\langle k \right|\left| p \right] = 0.
\end{split}
\end{equation}
Among the useful relations of the brackets one has the one that relates them directly to the four-vectors of the particles involved in the process of interest:
\begin{align}
\label{eq:brakets4vectores}
	\langle k\, p\rangle [p\, k] & = \mathbf{Tr}\left[\frac{1}{2}(1-\gamma_5)\slashed{k}\slashed{p}\right], \nonumber \\
    & = -2 k\cdot p\nonumber\\
    &= -(k+p)^2 .
\end{align}
(To get to the last line we've used the fact that the particles are massless.) We can write any ``massless'' (i.e., lightlike) slashed 4-vector $p$ in the ultrarrelativistic limit in the form
\begin{equation}
\label{eq:pslash}
-\slashed{p}=| p\rangle \left[p \right|+\left|p \right]\langle p| .
\end{equation}
\indent We will use the QED Feynman rules as usuallly; the difference with the \text{HF} is that our principal mathematical objects are now twistors, we should then express all objects appearing in the usual Feynman rules in terms of them. Until now we have expressed almost all the ingredients in terms of twistors, the only thing missing are the polarization four-vectors $\varepsilon_{\lambda_i}^{\mu *}$ and $\varepsilon_{\lambda'_i}^{\mu}$ ($\lambda_i$ represents the helicity) of the incoming and outgoing photons. \newline
\indent As seen in \cite{srednicki} the polarization vectors, in terms of twistors, can be written as follows:
\begin{align}
\label{eq:polarization}
\varepsilon_{+}^{\mu}(k) &=-\frac{\langle q|\gamma^{\mu}\left|k\right]}{\sqrt{2}\langle q\,k\rangle},\\
\varepsilon_{-}^{\mu}(k)&=-\frac{\left[ q\right|\gamma^{\mu}\left|k\right\rangle}{\sqrt{2}[ q\,k]}.
\end{align}
In the two previous expressions $q$ is an arbitrary massless reference vector. \newline
\indent It's convenient to have at hand the expressions for $\slashed{\varepsilon}_{\pm}(k)$, because they will appear in the expressions for the invariant amplitudes. We have
\begin{align}
\label{eq:polarizationslash}
\slashed{\varepsilon}_{+}(k;p) &=\frac{\sqrt{2}}{\langle q\,k\rangle}\left(\left|k\right]\langle q|+|q\rangle\left[k\right|\right),\\
\slashed{\varepsilon}_{-}(k;p) &=\frac{\sqrt{2}}{[q\,k]}\left(\left|k\right\rangle\left[ q\right|+\left|q\right]\langle k|\right).
\end{align}

\section{Compton scattering in the helicity formalism}
\label{sec:ComptonScatering}
The goal of this section is to calculate the invariant amplitude of Compton scattering $e^{-}\gamma \rightarrow e^{-}\gamma$ using the modern techniques of QFT such as the \textbf{HF}. In this work we will only do the analysis of Feynman diagrams at tree level. We define the Mandelstam variables:
\begin{equation}
\label{eq:mandelstam}
s_{ij}=-(p_i+p_j)^2.
\end{equation}

\begin{figure}[tbp]
\centering 
\includegraphics[width=10cm]{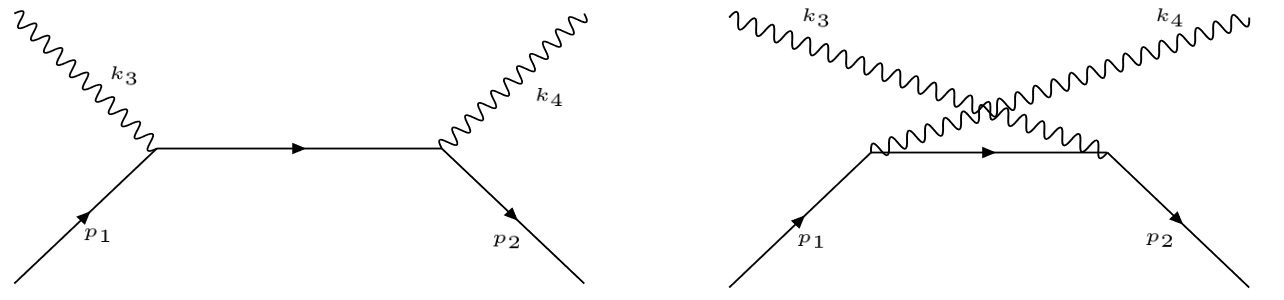}
\caption{\label{fig:compton} Diagrams for a fermion-photon scattering; we consider all the momenta to be outgoing.}
\end{figure}
It's possible to generalize the process $e^{-}\gamma \rightarrow e^{-}\gamma$ to interactions with two external fermions (not virtual) and two external photons, as shown in figure \ref{fig:compton}. It is important to identify the diagrams with helicity configurations that vanish; the diagrams that have two external fermions with the same helicity are zero, this is because the matrix elements of an odd number of gamma matrices in twistor space vanish \cite{griffiths_particles}; explicitly,
\begin{align}
\label{sandwishesgamma}
\langle p|\gamma^{\mu}|k \rangle&=0 ,\\
\left[p\right|\gamma^{\mu}\left| k\right]&=0 .
\end{align}
\indent We denote the amplitudes by $\mathcal{M}_{\lambda_1\lambda_2\lambda_3\lambda_4}$, where $\lambda_i$ is the helicity of the $i$-th particle. In our case, there are only two amplitudes that contribute to the process: $\mathcal{M}_{+-\lambda_3\lambda_4}$ and $\mathcal{M}_{-+\lambda_3\lambda_4}$, the later is the complex conjugate of the former, so we need only to calculate one of them. It's possible, starting from \eqref{eq:polarizationslash} and \eqref{eq:braketstwistors}, to show that
\begin{align}
\label{eq:condicionespolarizations}
\slashed{\varepsilon}_{-}(k;p)\left|p\right] &=0,\\
\label{eq:condicionespolarizations2}
\left[p\right|\slashed{\varepsilon}_{-}(k;p) &=0,\\
\label{eq:condicionespolarizations3}
\slashed{\varepsilon}_{+}(k;p)|p\rangle &=0,\\
\label{eq:condicionespolarizations4}
\langle p|\slashed{\varepsilon}_{+}(k;p) &=0.
\end{align}
The amplitude $\mathcal{M}_{+-\lambda_3\lambda_4}$ obtained directly from the conventional Feynman rules can be written in terms of twistors as follows
\begin{align}
\label{eq:amplitud1}
\mathcal{M}_{+-\lambda_3\lambda_4} & = (-i)e^2 \langle p_2|\varepsilon_{\lambda_4}^{\mu}(k_4;q_4)(i\gamma_{\mu})\left(\frac{-i(\slashed{p}_1
+\slashed{k}_3)}{(p_1+k_3)^2}\right)(i\gamma_{\nu})\varepsilon_{\lambda_3}^{\nu}(k_3;q_3)\left| p_1\right] \nonumber \\
 & \qquad {} -(-i)e^2\langle p_2|\varepsilon_{\lambda_3}^{\mu}(k_3;q_3)(i\gamma_{\mu})\left(\frac{-i(\slashed{p}_1
 + \slashed{k}_4)}{(p_1+k_4)^2}\right)(i\gamma_{\nu})\varepsilon_{\lambda_3}^{\nu}(k_4;q_4)\left| p_1 \right]\\
  \label{eq:amplitud2}
  & = -e^2 \langle 2|\slashed{\varepsilon}_{\lambda_4}(k_4;q_4)\left(\frac{\slashed{p_1}
 +\slashed{k_3}}{s_{13}}\right)\slashed{\varepsilon}_{\lambda_3}(k_3;q_3)\left| 1\right]\nonumber \\
 & \qquad {} -e^2\langle 2|\slashed{\varepsilon}_{\lambda_3}(k_3;q_3)\left(\frac{\slashed{p}_1
 + \slashed{k}_4}{s_{14}}\right)\slashed{\varepsilon}_{\lambda_3}(k_4;q_4)\left| 1 \right]
  \end{align}
If we take the polarizations $\lambda_3=\lambda_4=-$ in \eqref{eq:amplitud1} and we choose $q_3=q_4=p_1$ (remember that $q_3$ and $q_4$ are arbitrary reference 4-vectors) we get that $\mathcal{M}_{+-\lambda_3\lambda_4}$, after using \eqref{eq:condicionespolarizations}, is zero; the same occurs if $\lambda_3=\lambda_4=+$, choosing $q_3=q_4=p_2$ and using \eqref{eq:condicionespolarizations4}. The only amplitudes that survive are $\mathcal{M}_{+--+}$ and $\mathcal{M}_{+-+-}$. We see in \eqref{eq:amplitud2} that in the case $\lambda_3=+$, choosing the arbitrary reference four momentum $q_3=p_2$ and using \eqref{eq:condicionespolarizations4}, the second term of \eqref{eq:amplitud2} vanishes.
\begin{align}
\label{eq:amplitud5}
\mathcal{M}_{+-+-} &=-e^2 \langle 2|\slashed{\varepsilon}_{\lambda_4}(k_4;q_4)\left(\frac{\slashed{p}_1
 +\slashed{k}_3}{s_{13}}\right)\slashed{\varepsilon}_{\lambda_3}(k_3;q_3)\left| 1\right]
 \end{align}
Using \eqref{eq:polarizationslash} on \eqref{eq:amplitud5} we obtain:
 \begin{align}
\label{eq:amplitud6}
\mathcal{M}_{+-+-} &=-e^2 \left(\frac{1}{s_{13}}\right)\frac{\sqrt{2}}{[q_4\, 4]\langle 2 4\rangle}\left[q_4\right|(\slashed{p}_1
 +\slashed{k}_3)|2\rangle[3\,1]\frac{\sqrt{2}}{\langle 2\,3\rangle}
 \end{align}
Taking $q_4=k_3$ and remembering that $\left[p \right|\slashed{p}=\left[p \right|(| p\rangle \left[p\right|+\left|p\right]\langle p|)=0$, \eqref{eq:amplitud6} turns out to be
 \begin{align}
\label{eq:amplitud7}
\mathcal{M}_{+-+-} &=2e^2\frac{\langle 2\,4\rangle [3\,1] \langle 1\,2\rangle [3\,1]}{[3\,4] \langle 2\,3\rangle s_{13}} .
 \end{align}
\indent For a process with $n$ external particles and trying all the momenta as outgoing, momentum conservation is expressed as
  \begin{equation}
\label{eq:nparticulas}
\begin{split}
\sum_{j=1}^{n}\langle i\,j\rangle[j\,k] & = 0,\hspace{0.5cm} y
\qquad
\sum_{j=1}^{n}[i\,j]\langle j\,k\rangle  = 0 .
\end{split}
\end{equation}
From \eqref{eq:nparticulas} we can expand the sum for $n=4$, which is just the case for the process $e^{-}\gamma \rightarrow e^{-}\gamma$, and find that $\langle 2\,1 \rangle[1\,3]+\langle 2\,4 \rangle[4\,3]=0$; furthermore, we know that $s_{13}=\langle 1\,3\rangle [3\,1]$, considering this results in \eqref{eq:amplitud6} and cancelling equal terms, we obtain finally
 \begin{align}
\label{eq:amplitud7}
\mathcal{M}_{+-+-} &=2e^2\frac{\langle 2\,4\rangle^2}{\langle 1\,3\rangle \langle 2\,3\rangle} .
 \end{align}
The remaining amplitude $\mathcal{M}_{+--+}$ is found by crossing symmetry $3 \leftrightarrow 4$:
 \begin{align}
\label{eq:amplitud8}
\mathcal{M}_{+--+} &=2e^2\frac{\langle 2\,3\rangle^2}{\langle 1\,4\rangle \langle 2\,4\rangle} .
 \end{align}
The averaged squared amplitude is obtained in the usual way:
 \begin{align}
 \label{eq:amplitudtotal}
 \langle|\mathcal{M}|^2\rangle &=\frac{1}{4}\left[2\left(|\mathcal{M}_{+-+-}|^2+|\mathcal{M}_{+--+}|^2\right)\right] .
 \end{align}
Explicitly,
 \begin{align}
 \label{eq:amplitudtotalfinal1}
 \langle|\mathcal{M}|^2\rangle &=\frac{1}{4}\left\lbrace 2\left[4 e^4 \left(\frac{\langle2\,4\rangle^2}{\langle1\,3\rangle \langle2\,3\rangle}\right)\left(\frac{\langle2\,4\rangle^2}{\langle 1\,3\rangle \langle2\,3\rangle}\right)^{\ast}+4 e^4 \left(\frac{\langle2\,3\rangle^2}{\langle1\,4\rangle 2\,4}\right)\left(\frac{\langle2\,3\rangle^2}{\langle1\,4\rangle \langle2\,4\rangle}\right)^{\ast}\right]\right\rbrace .
 \end{align}
Taking the first term of \eqref{eq:amplitudtotalfinal1} and remembering that $[i\,j]^{\ast}=\langle j\,i \rangle$ we obtain, in the denominator, expressions of the type
\begin{align}
\langle1\,3\rangle\langle2\,3\rangle\langle1\,3\rangle^{\ast}\langle2\,3\rangle^{\ast} &= \langle1\,3\rangle\langle2\,3\rangle [1\,3][2\,3]\nonumber \\ &=s_{13}s_{23}
\end{align}
and, in the numerator,
\begin{align}
\langle2\,4\rangle\langle2\,4\rangle\langle2\,4\rangle^{\ast}\langle2\,4\rangle^{\ast} &= \langle2\,4\rangle\langle2\,4\rangle [2\,4][2\,4]\nonumber \\ &=s_{24}^2=s_{13}^2 .
\end{align}
In the same way we can simplify the second term of \eqref{eq:amplitudtotalfinal1} and, finally, obtain
\begin{align}
 \label{eq:amplitudtotalfinal2}
 \langle|\mathcal{M}|^2\rangle &=2 e^4\left(\left| \frac{s_{14}}{s_{13}} \right|+\left| \frac{s_{13}}{s_{14}} \right|\right)\nonumber\\
 &=-2e^4\left(\frac{u}{s}+\frac{s}{u}\right) .
 \end{align}
For Compton scattering, $s_{13}=s$, $s_{12}=t$ and $s_{14}=u$. We see that we obtain the same result as in the usual 4-component formalism, but we didn't had to use the algebra of traces, etc.

\section{Conclusions}
In this article we've presented the formalism for the treating fermions using Weyl spinors as the basic fields. Currently important developments to calculate dispersion processes with many particles in the final state, which are based on the called helicity formalism, are under progress. We consider it important to transmit the central features of these new techniques to the theoretical physics students, through the development of several exercises, including a physical scattering process. Among the examples discussed in this article, we include
\begin{enumerate}
\item A derivation of Weyl equations starting from K-G equation.
\item The incorporation of a fermion's magnetic moment interaction with the electromagnetic field $A_\mu$, through the principle of minimal substitution, i.e., $\partial_\mu\rightarrow D_\mu=\partial_\mu-ieA_\mu$.
\item The calculation of the amplitude for Compton scattering $e^-\gamma\rightarrow e^-\gamma$ using the helicity formalism, whose basic features were discussed at a basic level.
\end{enumerate}
With this paper we hope to achieve the goal of motivating the students to deepen their study of QFT and, in particular, in the subjects of the new helicity and amplitude methods.

\vspace{0.3in}
\large{\textbf{Acknowledgements}} \\
\normalsize
J. L. Diaz-Cruz acknowledges the support of SNI-CONACYT. B. Larios acknowledges the support of SMF and ICTP.

\end{document}